\documentclass[12pt]{article}
\pdfoutput=1

\renewcommand{\arraystretch}{1.2}
\usepackage{amsmath}
\usepackage{amssymb}
\usepackage{epsfig}
\usepackage{epstopdf}
\usepackage{latexsym}
\usepackage{color}
\numberwithin{equation}{section}
\usepackage[
      colorlinks=false,
      linkcolor=darkblue,  
      urlcolor=blue,    
      filecolor=blue,     
      citecolor=red,
linktocpage=true,
      pdfstartview=FitV,
      bookmarksopen=true    
      ]{hyperref}

\begin{document}

\begin{titlepage}

\centerline{\Huge \rm Supersymmetric Janus solutions} 
\bigskip
\centerline{\Huge \rm of dyonic $ISO(7)$-gauged $\mathcal{N}\,=\,8$ supergravity}
\bigskip
\bigskip
\bigskip
\bigskip
\bigskip
\bigskip
\centerline{\rm Minwoo Suh}
\bigskip
\centerline{\it Department of Physics, Kyungpook National University, Daegu 41566, Korea}
\bigskip
\centerline{\tt minwoosuh1@gmail.com} 
\bigskip
\bigskip
\bigskip
\bigskip
\bigskip
\bigskip
\bigskip

\begin{abstract}
\noindent We study supersymmetric Janus solutions of dyonic $ISO(7)$-gauged $\mathcal{N}$ = 8 supergravity. We mostly find Janus solutions flowing to 3d $\mathcal{N}$ = 8 SYM phase which is the worldvolume theory on D2-branes and non-conformal. There are also solutions flowing from the critical points which are dual to 3d SCFTs from deformations of the D2-brane theory.
\end{abstract}

\vskip 7cm

\flushleft {March, 2018}

\end{titlepage}

\tableofcontents

\section{Introduction}

Janus solutions provide simple and useful examples of the AdS/CFT correspondence \cite{Maldacena:1997re}. The Janus solutions are codimension one defects in supergravity, and they correspond to an interface in dual field theory. The Janus field theories and their gravity duals are most well studied in $\mathcal{N}$ = 4 super Yang-Mills theory.  The Janus field theories from $\mathcal{N}$ = 4 SYM were completely classified and constructed explicitly in \cite{Clark:2004sb, D'Hoker:2006uv, Gaiotto:2008sd}. The Janus solutions were constructed in type IIB supergravity in \cite{Bak:2003jk, D'Hoker:2006uu, D'Hoker:2007xy, D'Hoker:2007xz} or in gauged $\mathcal{N}$ = 8 supergravity in five dimensions in \cite{Clark:2005te} and uplifted to type IIB supergravity in \cite{Suh:2011xc}.

Although Janus field theories from ABJM theory are not known, the Janus solutions are proposed in eleven-dimensional supergravity \cite{DHoker:2008lup, DHoker:2008rje, DHoker:2009lky, DHoker:2009wlx}. The ABJM Janus solutions are also studied in gauged $\mathcal{N}$ = 8 supergravity in four dimensions, and some of them are uplifted to eleven-dimensions \cite{Bobev:2013yra, Pilch:2015dwa}. There are more examples of Janus solutions in four-dimensional gauged supergravity \cite{Karndumri:2016tpf, Karndumri:2016jls, Karndumri:2017bqi}. Lately, Janus solution was studied in $F(4)$ gauged supergravity in six dimensions, and was proposed to be dual to codimension one defect in 5d superconformal field theories \cite{Gutperle:2017nwo}.

In this paper, we study Janus solutions dual to 3d SCFTs from deformations of D2-brane theory via dyonic $ISO(7)$-gauged $\mathcal{N}$ = 8 supergravity. The well-known examples of the AdS/CFT correspondence on D3-, M2- and M5-branes involve anti-de Sitter spacetime as a near horizon geometry of the corresponding branes \cite{Maldacena:1997re}. There are  corresponding large N conformal field theories dual to the $AdS$ geometries, and they are $\mathcal{N}$ = 4 super Yang-Mills theory, ABJM theory and 6d (2,0) theory, respectively. When it comes to D2-branes, the near horizon geometry of a stack of D2-branes is not an $AdS$ spacetime, and its dual gauge theory is 3d $\mathcal{N}$ = 8 supersymmetric Yang-Mills theory which is non-conformal. However, by adding a Chern-Simons term, 3d $\mathcal{N}$ = 8 SYM flows to a conformal fixed point with $\mathcal{N}$ = 2, $U(1)_R{\times}SU(3)$ symmetry \cite{Schwarz:2004yj, Gaiotto:2007qi}. Otherwise, by adding a mass term to one of the chiral scalars, it flows to another conformal fixed point with $\mathcal{N}$ = 3, $SO(4)$ symmetry \cite{Minwalla:2011ma}. Recently, gravity duals of these 3d SCFTs from D2-branes were discovered in dyonic $ISO(7)$-gauged $\mathcal{N}$ = 8 supergravity \cite{Guarino:2015jca}.

Dyonic $ISO(7)$-gauged $\mathcal{N}$ = 8 supergravity is a consistent truncation of massive type IIA supergravity on six-sphere \cite{Guarino:2015qaa, Guarino:2015vca, Varela:2015uca}. The Romans mass corresponds to the magnetic gauge coupling of dyonic $ISO(7)$-gauged $\mathcal{N}$ = 8 supergravity. The scalar potential of the theory has four known superymmetric critical points, and they are $\mathcal{N}$ = 2 $SU(3){\times}U(1)$, $\mathcal{N}$ = 1 $SU(3)$, $\mathcal{N}$ = 1 $G_2$, and $\mathcal{N}$ = 3 $SO(4)$ critical points. The $\mathcal{N}$ = 2 $SU(3){\times}U(1)$ and $\mathcal{N}$ = 3 $SO(4)$ critical points are dual to the 3d SCFTs from the deformations of D2-brane theory discussed in the previous paragraph. The fixed point $AdS$ solutions were uplifted to massive type IIA supergravity \cite{Guarino:2015vca}, and holographic RG flows between critical points were studied \cite{Guarino:2016ynd}. The gravitational free energies \cite{Guarino:2015qaa, Fluder:2015eoa, Pang:2015vna} and spin-2 spectrum \cite{Pang:2015rwd, Pang:2017omp} were calculated and matched with the field theory results.

In the usual supergravity theories, most of the RG flows and Janus solutions are attracted to critical points which are dual to conformal field theories $e.g.$ $\mathcal{N}$ = 4 super Yang-Mills theory, ABJM theory, and 6d (2,0) theory. On the other hand, in dyonic $ISO(7)$-gauged $\mathcal{N}$ = 8 supergravity, most of the RG flows and Janus solutions are attracted to a non-conformal phase which is dual to 3d $\mathcal{N}$ = 8 SYM on the worldvolume of D2-branes. Therefore, as we will see, usual Janus solutions in dyonic $ISO(7)$-gauged $\mathcal{N}$ = 8 supergravity do not exhibit $AdS$-behavior. On the other hand, if we fine-tune initial values of the scalar fields, we obtain Janus solutions staying at a critical point for a while and then moving away to the non-conformal 3d SYM phase.

In section 2, we review the $\mathcal{N}$ = 1, $\mathbb{Z}_2{\times}SO(3)$-invariant truncation of dyonic $ISO(7)$-gauged $\mathcal{N}\,=\,8$ supergravity. In section 3, we study Janus solutions in the the $\mathcal{N}$ = 2, $SU(3){\times}U(1)$-invariant truncation. In section 4, we study Janus solutions in the the $\mathcal{N}$ = 1, $G_2$-invariant truncation. We conclude in section 5. In appendix A we present the equations of motion from the truncations we consider. In appendix B, we present the $\mathcal{N}$ = 1, $SO(4)$-invariant truncation.

\section{Dyonic $ISO(7)$-gauged $\mathcal{N}\,=\,8$ supergravity}

We begin by considering the $\mathcal{N}$ = 1, $\mathbb{Z}_2{\times}SO(3)$-invariant truncation of dyonic $ISO(7)$-gauged $\mathcal{N}\,=\,8$ supergravity. This truncation was studied in detail in appendix A of \cite{Guarino:2015qaa}, and we review it in this section. There are three complex scalar fields, $\Phi_1$, $\Phi_2$ and $\Phi_3$. In terms of canonical $\mathcal{N}\,=\,1$ formulation, the scalar action is given by
\begin{equation}
S\,=\,\frac{1}{16{\pi}G_4}{\int}d^4x\sqrt{-g}\left[R+6\frac{\partial_\mu\Phi_1\partial^\mu\overline{\Phi}_1}{(\Phi_1-\overline{\Phi}_1)^2}+2\frac{\partial_\mu\Phi_2\partial^\mu\overline{\Phi}_2}{(\Phi_2-\overline{\Phi}_2)^2}+6\frac{\partial_\mu\Phi_3\partial^\mu\overline{\Phi}_3}{(\Phi_3-\overline{\Phi}_3)^2}-\mathcal{P}\,\right].
\end{equation}
The K\"ahler potential is
\begin{equation}
\mathcal{K}\,=\,-3\log\left[-i(\Phi_1-\overline{\Phi}_1)\right]-\log\left[-i(\Phi_2-\overline{\Phi}_2)\right]-3\log\left[-i(\Phi_3-\overline{\Phi}_3)\right]\,,
\end{equation}
and the holomorphic superpotential is 
\begin{equation}
\mathcal{V}\,=\,2g\left(\Phi_1^3+3\Phi_1\Phi_3^2+3\Phi_1\Phi_2\Phi_3\right)+2m\,.
\end{equation}
The K\"ahler metric is defined by
\begin{equation}
\mathcal{K}_{\Phi_p\bar{\Phi}_q}\,=\,\partial_{\Phi_p}\partial_{\bar{\Phi}_q}\mathcal{K}\,.
\end{equation}
Then, we define the complex superpotential,{\footnote{There is an additional factor 2 in the complex superpotential defined in (2.3) of \cite{Guarino:2016ynd} compared to (3.27) of \cite{Guarino:2015qaa}. Through the paper, we follow the normalization of (2.3) in \cite{Guarino:2016ynd}, as it gives standard normalization for the flow equations.}}
\begin{equation}
\mathcal{W}\,=\,\sqrt{2}e^{\mathcal{K}/2}\mathcal{V}\,,
\end{equation}
and the real superpotential,
\begin{equation}
W^2\,=\,|\mathcal{W}|^2\,.
\end{equation}
The scalar potential is obtained from
\begin{equation} \label{scp}
\mathcal{P}\,=\,2\left[-\frac{4}{3}(\Phi_1-\bar{\Phi}_1)^2\left|\frac{\partial{W}}{\partial{\Phi_1}}\right|^2-4(\Phi_2-\bar{\Phi}_2)^2\left|\frac{\partial{W}}{\partial{\Phi_2}}\right|^2-\frac{4}{3}(\Phi_3-\bar{\Phi}_3)^2\left|\frac{\partial{W}}{\partial{\Phi_3}}\right|^2-3W^2\right]\,.
\end{equation}
The scalar potential has all four known supersymmetric critical points of dyonic $ISO(7)$-gauged $\mathcal{N}$ = 8 supergravity \cite{Guarino:2015qaa}, and they are listed in Table 1. The ratio of electric and magnetic gauge couplings, $g$ and $m$, respectively, is denoted by $c$ = $m/g$. 

\begin{center}
\renewcommand{\arraystretch}{1}
\begin{tabular}{c|c|c} 
\cline{1-3} 
$\mathcal{N}$ \qquad \qquad\,\,\,\,\,\,\,\, $G_0$\,\,\,\,\,\,\,\, & $c^{-1/3}\chi$ \qquad $c^{-1/3}e^{-\varphi}$ \qquad $c^{-1/3}\rho$ \qquad $c^{-1/3}e^{-\phi}$ & $g^{-2}c^{-1/3}V_0$ \\ \cline{1-3}
$\mathcal{N}$ = 3 \qquad\,\,\,\,\,\,\,\, $SO(4)$\,\,\,\,\,\,\,\, & $\,\,\,\,\,\,\,\,\,\,\,\,\,\,\,\,\,\,\,\,\,\,\,\,\, \frac{1}{2^{4/3}} \qquad \qquad \frac{3^{1/2}}{2^{4/3}} \qquad \,\, -\frac{1}{2^{1/3}} \,\,\,\,\,\,\,\, \qquad \frac{3^{1/2}}{2^{1/3}} \,\,\,\,\,\,\,\,\,\,\,\,\,\,\,\,\,\,\,\,\,\,\,\,$ & $-\frac{2^{16/3}}{3^{1/2}}$ \\ \cline{1-3}
$\mathcal{N}$ = 2 \qquad $SU(3){\times}U(1)$ & $-\frac{1}{2} \qquad \qquad \frac{3^{1/2}}{2} \qquad \qquad 0 \qquad \qquad \frac{1}{2^{1/2}}$ & $-2^23^{3/2}$  \\ \cline{1-3}
$\mathcal{N}$ = 1 \qquad\,\,\,\,\,\,\,\,\,\,\,\, $G_2$\,\,\,\,\,\,\,\,\,\,\,\,\,\, & $-\frac{1}{2^{7/3}} \qquad \,\,\,\,\,\, \frac{5^{1/2}3^{1/2}}{2^{7/3}} \qquad -\frac{1}{2^{7/3}} \qquad \frac{5^{1/2}3^{1/2}}{2^{7/3}}$ & $-\frac{2^{28/3}3^{1/2}}{5^{5/2}}$  \\ \cline{1-3}
$\mathcal{N}$ = 1 \qquad\,\,\,\,\,\,\,\, $SU(3)$\,\,\,\,\,\,\, & $ \,\,\,\, \frac{1}{2^2} \qquad \,\,\,\,\,\,\,\,\, \frac{3^{1/2}5^{1/2}}{2^2} \qquad -\frac{3^{1/2}}{2^2} \qquad \,\,\,\,\,\, \frac{5^{1/2}}{2^2}$ & $-\frac{2^83^{3/2}}{5^{5/2}}$  \\ \cline{1-3}
\end{tabular} \\ 
\medskip
{\it Table 1. All known supersymmetric critical points of dyonic $ISO(7)$-gauged $\mathcal{N}$ = 8 supergravity.}
\end{center}

The $\mathcal{N}$ = 2 $SU(3)$-, $\mathcal{N}$ = 1 $SO(4)$-, and $\mathcal{N}$ = 1 $G_2$-invariant truncations are obtained as sub-truncations by identifying the complex scalar fields as, \cite{Guarino:2015qaa},
\begin{align} \label{sub}
&\text{SU(3) truncation:} \qquad \Phi_1\,=\,t\,, \qquad \Phi_2\,=\,\Phi_3\,=\,u\,, \notag \\
&\text{SO(4) truncation:} \qquad \Phi_1\,=\,\Phi_3\,=\,t\,, \qquad \Phi_2\,=\,u\,, \notag \\
&\text{$G_2$ truncation:} \qquad\,\,\,\,\,\,\,\,\, \Phi_1\,=\,\Phi_2\,=\,\Phi_3\,=\,t\,,
\end{align}
where we introduce a parametrization in real scalar fields,
\begin{equation}
t\,=\,-\chi+ie^{-\varphi}\,, \qquad u\,=\,-\rho+ie^{-\phi}\,.
\end{equation}
In addition to the $AdS$ critical points in Table 1, the non-conformal 3d $\mathcal{N}$ = 8 SYM phase is at $t,\,u\,\rightarrow\,+\,i\,\infty$, which corresponds to $\chi,\,\rho\,=\,0$ and $\varphi,\,\phi\,\rightarrow\,-\,\infty$, \cite{Guarino:2016ynd}. 

When obtaining a scalar potential of the sub-truncations, one has to first perform the differentiations in \eqref{scp}, and then identify the scalar fields by \eqref{sub}. Otherwise, one has to employ the scalar potential formula from the canonical $\mathcal{N}$ = 1 formulation, \cite{Guarino:2015qaa},
\begin{equation}
\mathcal{P}\,=\,e^\mathcal{K}\left[\mathcal{K}^{\Phi_p\bar{\Phi}_q}(D_{{\Phi}_p}\mathcal{V})(D_{{\bar{\Phi}}_p}\mathcal{\overline{V}})-3\mathcal{V}\overline{\mathcal{V}}\right]\,,
\end{equation}
where the K\"ahler covariant derivative is
\begin{equation}
D_{{\Phi}_p}\mathcal{V}\,=\,\partial_{{\Phi}_p}\mathcal{V}+(\partial_{{\Phi}_p}\mathcal{K})\mathcal{V}\,.
\end{equation}


\section{The $SU(3)$-invariant Janus solutions}

The scalar action for the $SU(3)$-invariant truncation, \cite{Guarino:2015qaa, Guarino:2016ynd}, is obtained by \eqref{sub},
\begin{equation} \label{su3act}
S\,=\,\frac{1}{16{\pi}G_4}{\int}d^4x\sqrt{-g}\left[R+6\frac{\partial_\mu{t}\partial^\mu\overline{t}}{(t-\overline{t})^2}+8\frac{\partial_\mu{u}\partial^\mu\overline{u}}{(u-\overline{u})^2}-\mathcal{P}\,\right].
\end{equation}
The K\"ahler potential is
\begin{equation}
\mathcal{K}\,=\,-3\log\left[-i(t-\overline{t})\right]-4\log\left[-i(u-\overline{u})\right]\,,
\end{equation}
and the holomorphic superpotential is
\begin{equation}
\mathcal{V}\,=\,2g\left(t^2+6tu^2\right)+2m\,.
\end{equation}
The scalar potential is obtained from
\begin{equation}
\mathcal{P}\,=\,2\left[-\frac{4}{3}(t-\bar{t})^2\left|\frac{\partial{W}}{\partial{t}}\right|^2-(u-\bar{u})^2\left|\frac{\partial{W}}{\partial{u}}\right|^2-3W^2\right]\,.
\end{equation}
The scalar potential has $\mathcal{N}$ = 2 $SU(3){\times}U(1)$-, $\mathcal{N}$ = 1 $SU(3)$-, and $\mathcal{N}$ = 1 $G_2$-invariant critical points, and more non-supersymmetric critical points.

For Janus solutions, we consider the $AdS_3$-sliced domain wall for background,
\begin{equation}
ds^2\,=\,e^{2A(r)}d^2_{AdS_3}+dr^2\,.
\end{equation}
Then, we solve the supersymmetry variations of fermionic fields on the curved background, as it was done on the Lorentzian $AdS_3$-domain walls in \cite{Bobev:2013yra} or on the Euclidean $S^3$-domain walls in \cite{Freedman:2013ryh}. The supersymmetry equations for the complex scalar fields,
\begin{align}
t'\,=&\,\left(-i \kappa\frac{e^{-A}}{l}-A'\right)\mathcal{K}^{t\bar{t}}\frac{2}{W}\frac{{\partial}W}{\partial\bar{t}}\,, \notag \\
u'\,=&\,\left(-i \kappa\frac{e^{-A}}{l}-A'\right)\mathcal{K}^{u\bar{u}}\frac{2}{W}\frac{{\partial}W}{\partial\bar{u}}\,,
\end{align}
and for the warp factor,
\begin{equation} \label{warp}
A'\,=\,\pm\sqrt{W^2-\frac{e^{-2A}}{l^2}}\,,
\end{equation}
are obtained where $l$ is the radius of $AdS$ and $\kappa\,=\,\pm1$. The constant, $\kappa\,=\,\pm1$, is not related to the signs in \eqref{warp}. Reversing the sign of $\kappa$ merely generates a solution reflected in the $r$-coordinate. In terms of the real scalar fields, the supersymmetry equations are
\begin{align} \label{su3susy}
\chi'+\frac{4}{3}e^{-2\varphi}\frac{A'}{W}\frac{{\partial}W}{\partial\chi}-\frac{4\kappa}{3}\frac{e^{-A-\varphi}}{l}\frac{1}{W}\frac{{\partial}W}{\partial\varphi}\,=&\,0\,, \notag \\
\varphi'+\frac{4}{3}\frac{A'}{W}\frac{{\partial}W}{\partial\varphi}+\frac{4\kappa}{3}\frac{e^{-A-\varphi}}{l}\frac{1}{W}\frac{{\partial}W}{\partial\chi}\,=&\,0\,, \notag \\
\rho'+e^{-2\phi}\frac{A'}{W}\frac{{\partial}W}{\partial\rho}-\kappa\frac{e^{-A-\phi}}{l}\frac{1}{W}\frac{{\partial}W}{\partial\phi}\,=&\,0\,, \notag \\
\phi'+\frac{A'}{W}\frac{{\partial}W}{\partial\phi}+\kappa\frac{e^{-A-\phi}}{l}\frac{1}{W}\frac{{\partial}W}{\partial\rho}\,=&\,0\,.
\end{align}

Now we numerically obtain supersymmetric Janus solutions of the $SU(3)$-invariant truncation.{\footnote{In $SO(8)$-gauged $\mathcal{N}$ = 8 supergravity, the $SU(3){\times}U(1){\times}U(1)$ sector was considered to study Janus solutions in \cite{Bobev:2013yra}. In dyonic $ISO(7)$-gauged $\mathcal{N}$ = 8 supergravity, one can turn off the scalar field, $\rho$, and obtain the $SU(3){\times}U(1)$-invariant truncation \cite{Guarino:2015jca, Guarino:2015qaa}. However, it is not consistent to turn off the scalar field, $\phi$, and there is no $SU(3){\times}U(1){\times}U(1)$-invariant truncation. One can see this by examining the equation of motion for $\phi$. Moreover, it is group theoretically impossible as $ISO(7)$ does not have an $SU(3){\times}U(1){\times}U(1)$ subgroup.}} We closely follow the method employed in \cite{Bobev:2013yra}. We set $l\,=\,1$, $g\,=\,1$, $m\,=\,1$, and $\kappa\,=\,+1$ for all numerical solutions in this paper. As the supersymmetry equation for the warp factor, \eqref{warp}, involves a square root, it involves a branch cut. In order to obtain smooth Janus solutions, we should choose the upper sign for $r>0$ and lower sign for $r<0$, \cite{Clark:2005te, Suh:2011xc, Bobev:2013yra, Gutperle:2017nwo}. In order to avoid dealing with branch cuts, instead of the first-order supersymmetry equations, we solve the second-order equations of motion which are presented in appendix A. To numerically solve the second-order equations of motion, we should specify the initial conditions for the scalar fields, $\varphi$, $\chi$, $\phi$, $\rho$, the warp factor, $A$, and their derivatives. We impose a smoothness condition at the interface,
\begin{equation} \label{smooth}
A'(0)\,=\,0.
\end{equation}
Because of the smoothness condition, if we fix initial conditions for the scalar fields, \{$\varphi(0)$, $\chi(0)$, $\phi(0)$, $\rho(0)$\}, initial condition for the warp factor, $A(0)$ is automatically determined through the supersymmetry equation for $A(r)$ in \eqref{warp}. Similarly, \{$\varphi'(0)$, $\chi'(0)$, $\phi'(0)$, $\rho'(0)$\} are also determined through the supersymmetry equations for the scalar fields in \eqref{su3susy}. Finally, we have only four free parameters, \{$\varphi(0)$, $\chi(0)$, $\phi(0)$, $\rho(0)$\}, and when they are fixed, the rest of the parameters are fixed with the smoothness condition, \eqref{smooth}, and the supersymmetry equations. When we obtain numerical solutions of the equations of motion, we numerically check whether they satisfy the supersymmetry equations.

As we will see in detail, we find that there are three classes of solutions: regular Janus solutions, singular solutions, and as a special case of regular Janus solutions, there are Janus solutions flowing from a critical point. The solutions are effectively depicted on contour plot of the superpotential. Following \cite{Bobev:2009ms}, in order to have both $\mathcal{N}$ = 2, $SU(3){\times}U(1)$ and $\mathcal{N}$ = 1, $SU(3)$ critical points depicted on contour plots, we replace some scalar fields by
\begin{equation} \label{pr}
\phi\,=\,\phi_2+(\phi_1-\phi_2)\frac{\varphi^2-\varphi_2^2}{\varphi_1^2-\varphi_2^2}\,, \qquad \rho\,=\,\rho_2+(\rho_1-\rho_2)\frac{\varphi^2-\varphi_2^2}{\varphi_1^2-\varphi_2^2}\,,
\end{equation}
or
\begin{equation} \label{pp}
\phi\,=\,\phi_2+(\phi_1-\phi_2)\frac{\chi^2-\chi_2^2}{\chi_1^2-\chi_2^2}\,, \qquad \varphi\,=\,\varphi_2+(\varphi_1-\varphi_2)\frac{\chi^2-\varphi_2^2}{\chi_1^2-\chi_2^2}\,,
\end{equation}
where $\varphi_2$, $\chi_2$, $\phi_2$ are the values of the scalar fields at the $\mathcal{N}$ = 2, $SU(3)$ critical point, and $\varphi_1$, $\chi_1$, $\phi_1$ at the $\mathcal{N}$ = 1, $G_2$ critical point.
In Figure 1, the contour plot with \eqref{pr} is presented on the left  and \eqref{pp} on the right. Regular, singular and critical-point Janus solutions are depicted by blue, green and red lines, respectively.

\begin{figure}[h!]
\begin{center}
\includegraphics[width=3.0in]{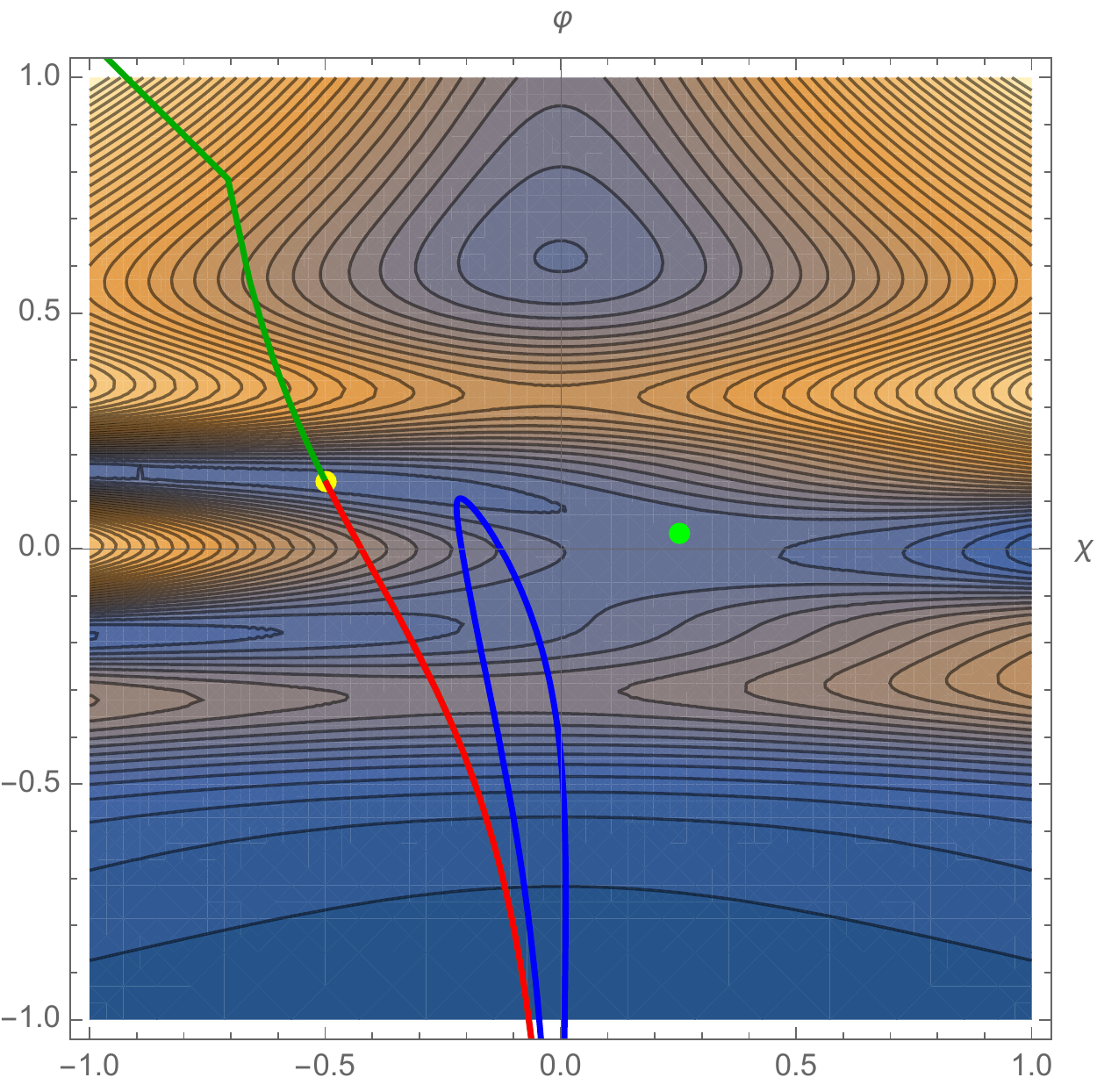} \qquad \includegraphics[width=3.0in]{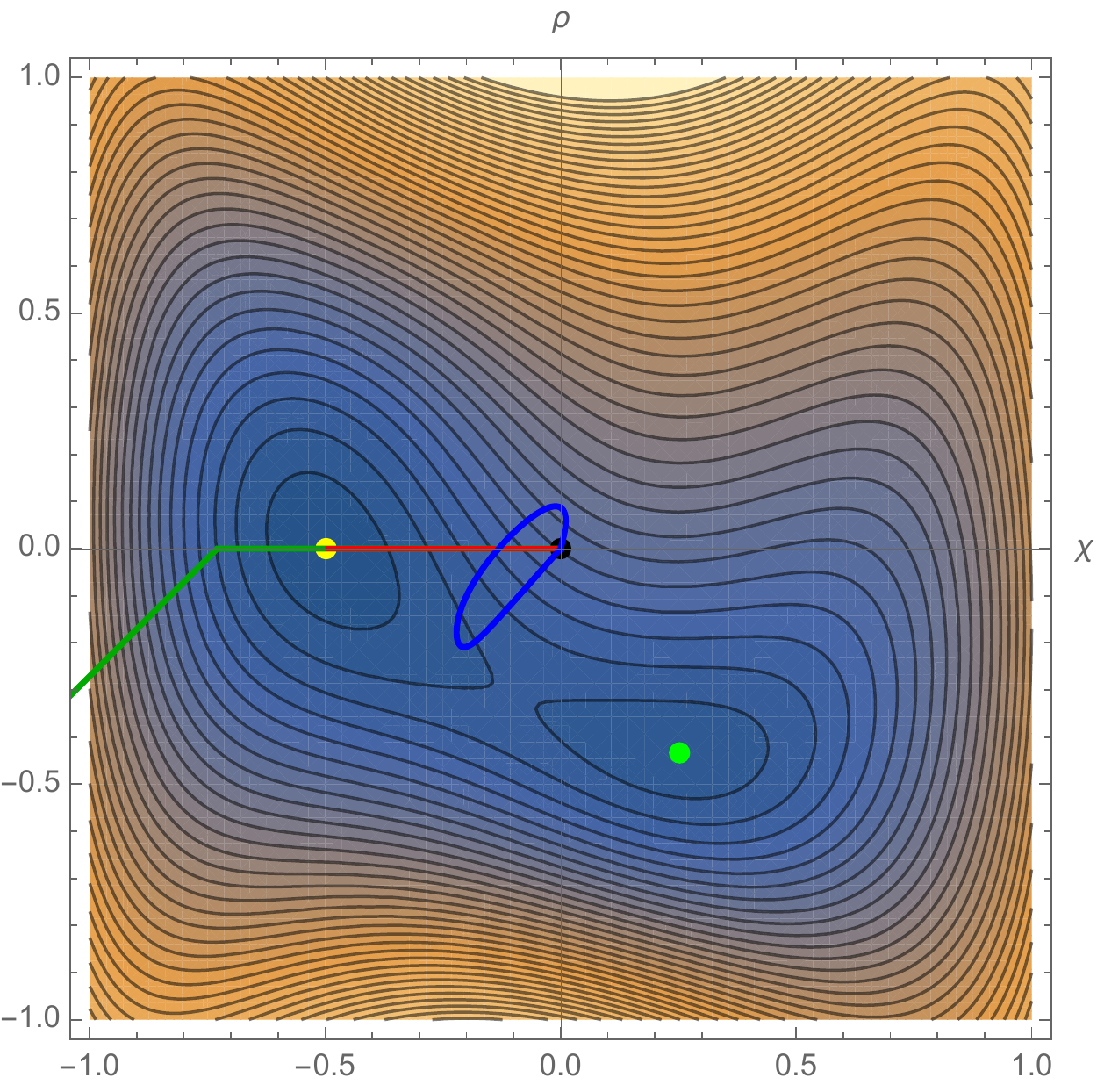}
\caption{{\it Contour plot of the $SU(3)$-invariant superpotential and flows. The $\mathcal{N}$ = 2, $SU(3){\times}U(1)$ and $\mathcal{N}$ = 1, $SU(3)$ critical points are denoted by yellow and green dots, respectively. In the left plot, the 3d $\mathcal{N}$ = 8 SYM phase is located at $\chi\,=\,0$ and $\varphi\,\rightarrow\,-\,\infty$. In the right plot, the 3d $\mathcal{N}$ = 8 SYM phase is at the origin, and is denoted by a black dot.}}
\label{1}
\end{center}
\end{figure}

\begin{figure}[h!]
\begin{center}
\includegraphics[width=2.0in]{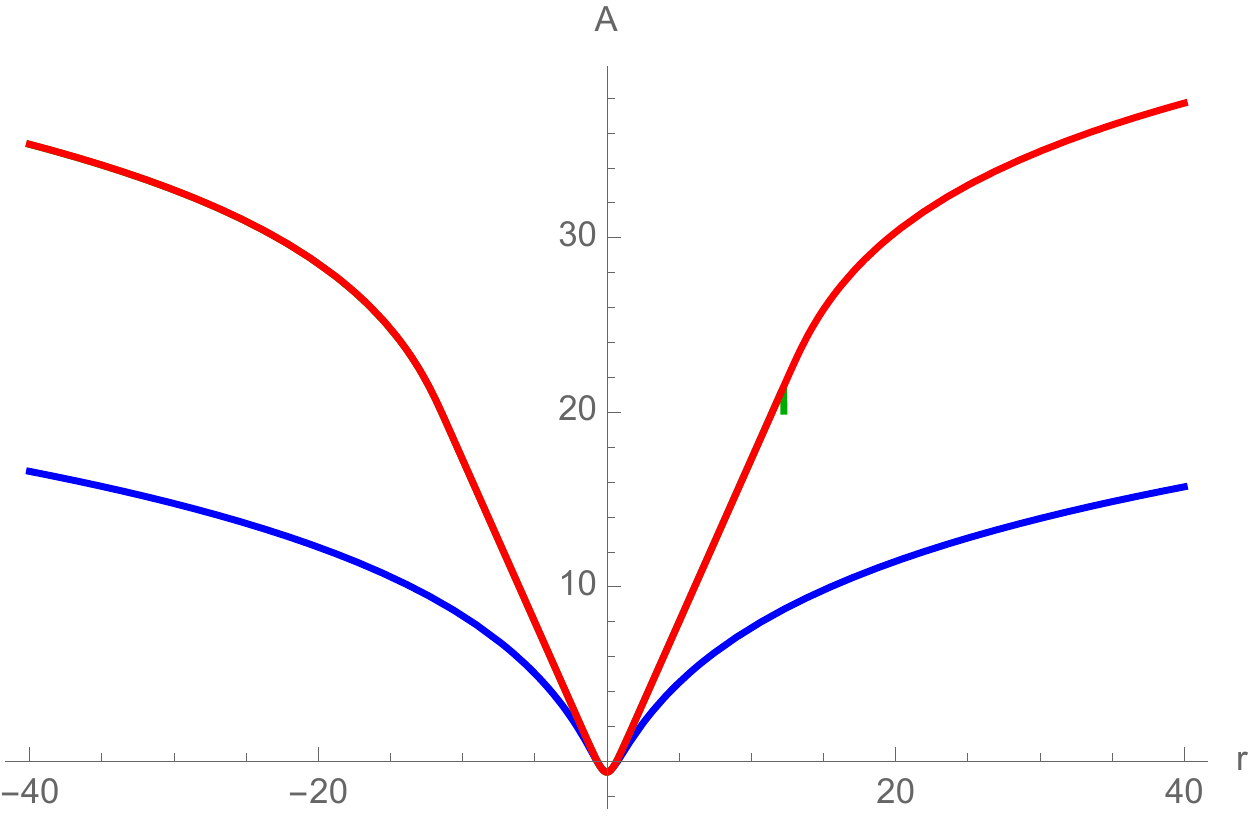} \qquad \includegraphics[width=2.0in]{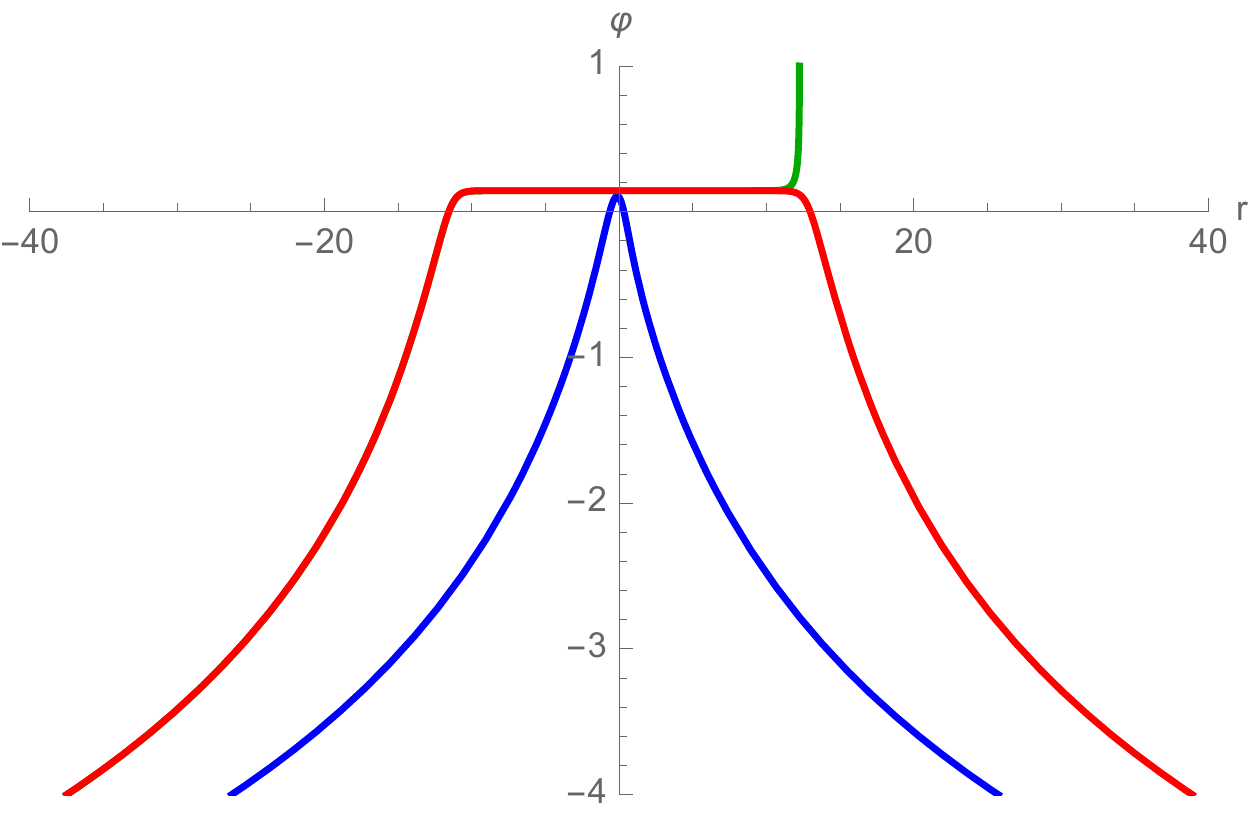} \qquad \includegraphics[width=2.0in]{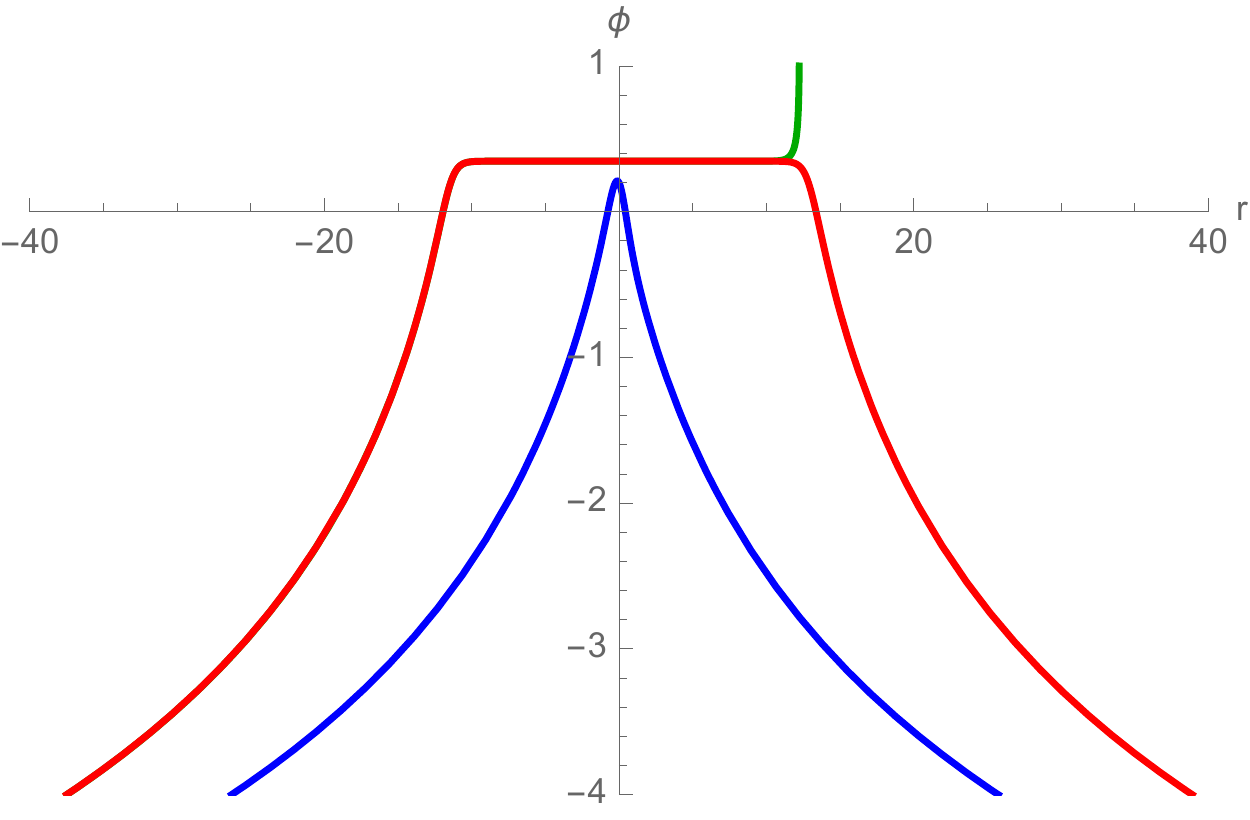} \\
\includegraphics[width=2.0in]{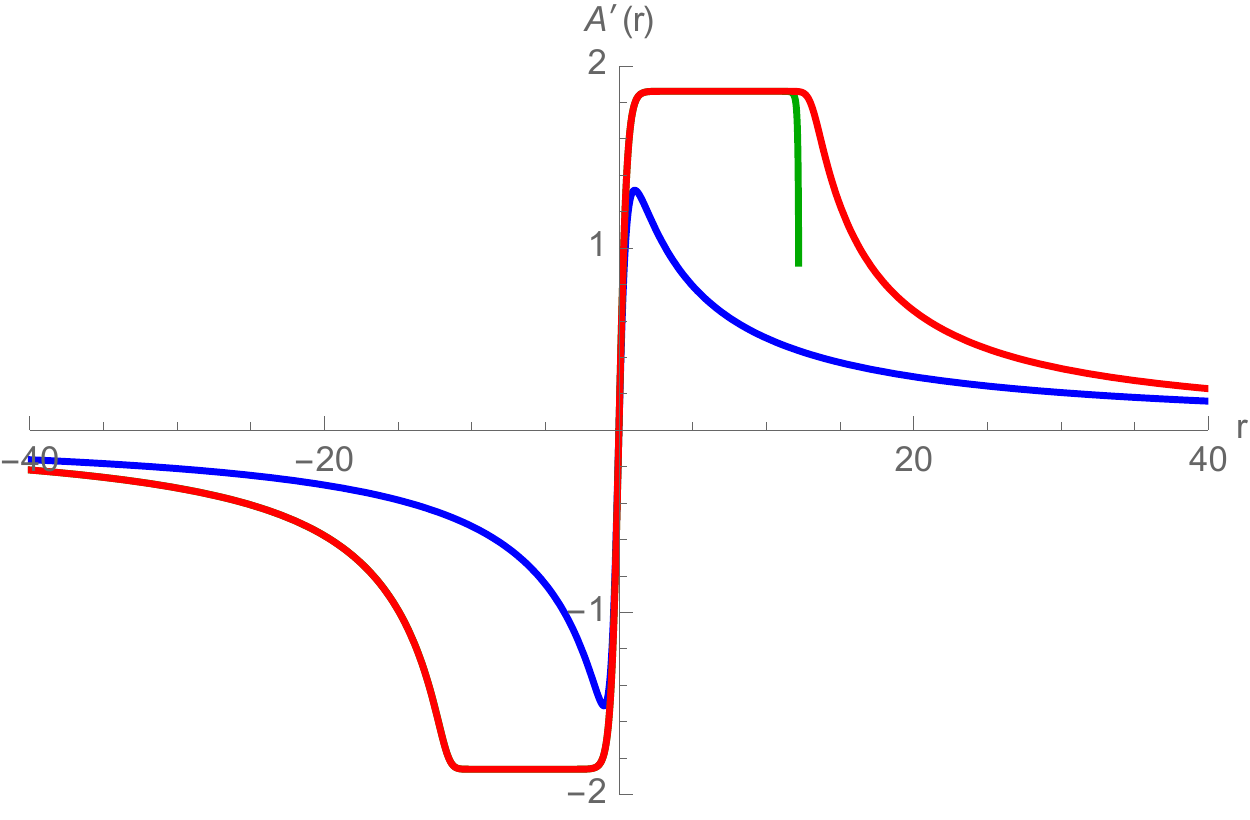} \qquad \includegraphics[width=2.0in]{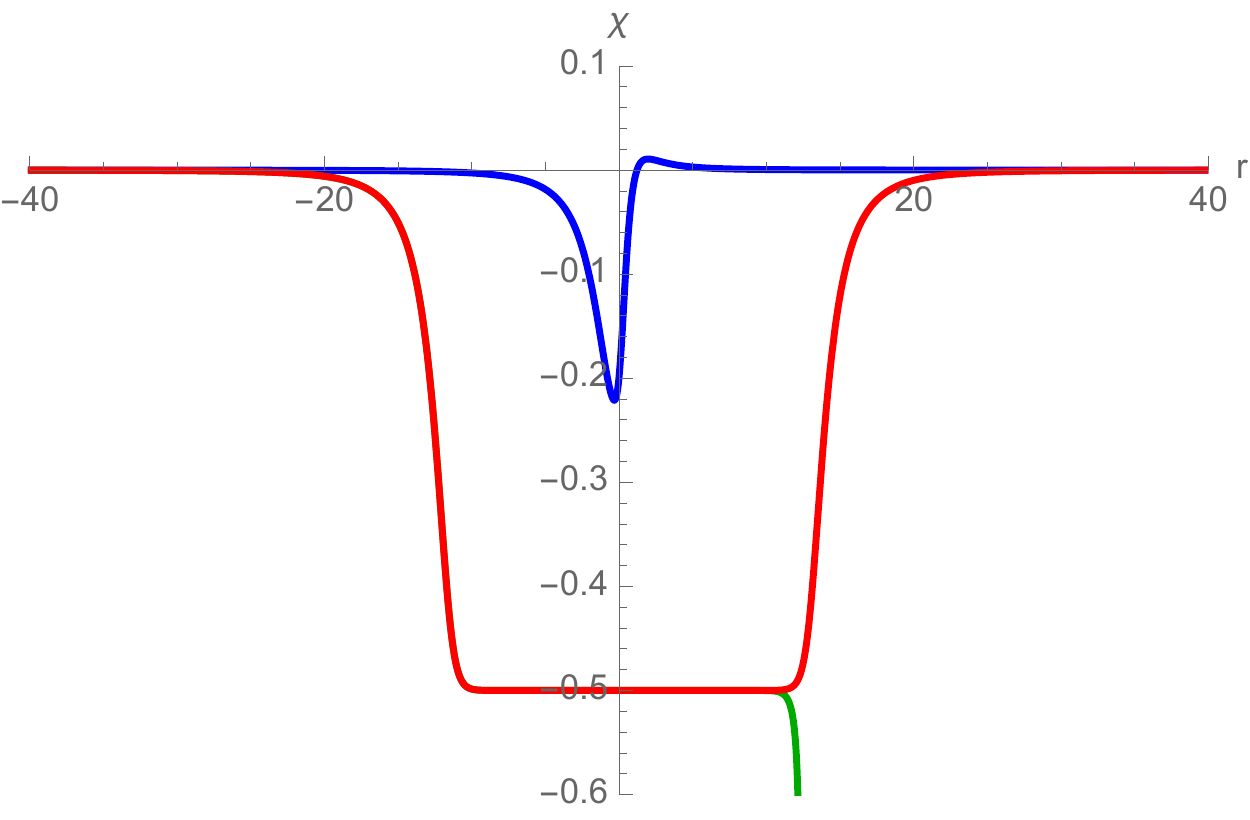} \qquad \includegraphics[width=2.0in]{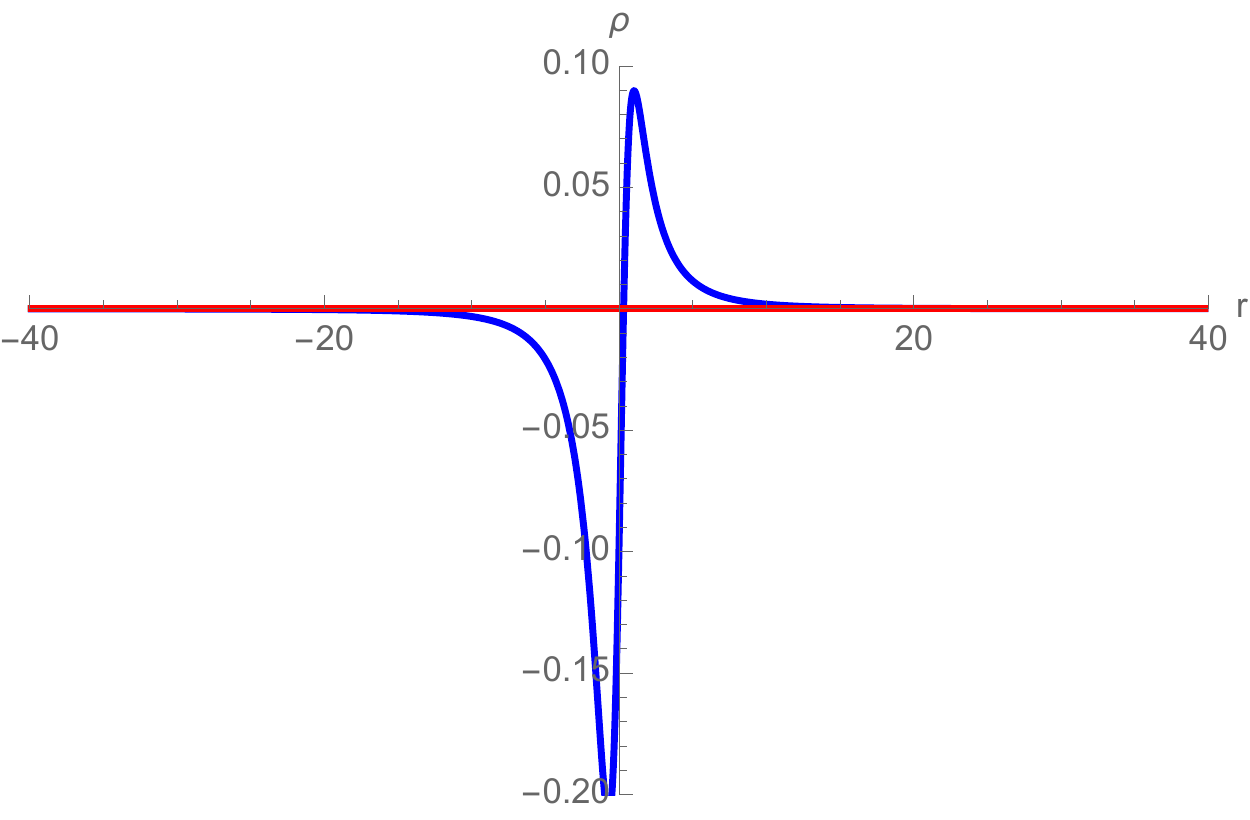}
\caption{{\it Three solutions with \{$\varphi(0)\,=\,0.1$, $\chi(0)\,=\,-0.2$, $\phi(0)\,=\,0.2$, $\rho(0)\,=\,-0.1$\} (Blue), \{$\varphi(0)\,=\,\varphi_*+2.9\times10^{-16}$, $\chi(0)\,=\,\chi_*$, $\phi(0)\,=\,\phi_*$, $\rho(0)\,=\,\rho_*$\} (Red), and \{$\varphi(0)\,=\,\varphi_*+3.0\times10^{-16}$, $\chi(0)\,=\,\chi_*$, $\phi(0)\,=\,\phi_*$, $\rho(0)\,=\,\rho_*$\} (Green).}}
\label{1}
\end{center}
\end{figure}

First class of solutions are regular Janus solutions. Typical initial conditions give solutions attracted to 3d $\mathcal{N}$ = 8 SYM phase which is non-conformal.{\footnote {This is the main difference between the solutions here and the known Janus solutions of four- and five-dimensional gauged supergravity where the solutions are mostly attracted to the $conformal$ points which are dual to 3d ABJM \cite{Bobev:2013yra} and 4d $\mathcal{N}$=4 SYM \cite{Clark:2005te, Suh:2011xc}, respectively. They show constant slope behavior of $AdS$ solutions all along the flow.}} Therefore, warp factors of the solutions do not show $AdS$-behavior of constant slope. To the best of our knowledge, they are the first examples of $non$-$conformal$ Janus solutions. We plot a typical regular Janus solution by blue lines in Figures 1 and 2.

The more interesting class of solutions arises as we approach the critical point. As we have four scalar fields, we fix initial conditions of three scalar fields, $\chi(0)\,=\,\chi_*$, $\phi(0)\,=\,\phi_*$, $\rho(0)\,=\,\rho_*$, by their critical point values, and we have only one initial condition to fix, $\varphi(0)$. When $\varphi(0)<\varphi_{cr}$, we obtain the first class of solutions, $i.e.$ regular Janus solutions. We found the value of $\varphi_{cr}$ to be
\begin{equation}
\varphi_*+1.6\times10^{-16}\,\leq\,\varphi_{cr}\,\leq\,\varphi_*+2.9\times10^{-16}\,.
\end{equation}
When we fine-tune $\varphi(0)=\varphi_{cr}$, numerical solution starts approaching the critical point, stays at the critical point for a while, and then move away to the 3d SYM phase. While the solution is staying at the critical point, the warp factor exhibits the $AdS$-behavior with a constant slope. The warp factor is given by
\begin{equation}
A(r)\,\sim\,W_*r,
\end{equation}
where the slope, $W_*$, is the value of the superpotential at the critical point. We plot a typical critical-point solution by red lines in Figures 1 and 2. On the contour plot in Figure 1, this solution takes the steepest descent from the $SU(3){\times}U(1)$ critical point to the SYM phase.{\footnote {The three classes of solutions we found are very much analogous to the solutions found in $SO(8)$-gauged $\mathcal{N} \,=\,8$ supergravity in Figures 7 and 8 of \cite{Bobev:2013yra}. Unlike our case, in $SO(8)$-gauged $\mathcal{N}$ = 8 supergravity, the critical points exist in pairs: they are $\mathbb{Z}_2$-symmetric to each other.}}

Regularity of the solution flowing from the $SU(3){\times}U(1)$ critical point can be seen from two aspects. First, as one can see from Figure 2 where the solution is plotted in red, the numerical solution is well-defined though all range of $r$. The singular solution we are about to present is not well-defined at some finite $r$ and diverges. Second, on the contour plot of the superpotential in Figure 1, the point dual to 3d $\mathcal{N}$ = 8 SYM phase is located at $\chi\,=\,0$ and $\varphi\,\rightarrow\,-\,\infty$ \cite{Guarino:2016ynd}{\footnote {In the parametrization of \cite{Guarino:2016ynd}, the 3d $\mathcal{N}$ = 8 SYM phase corresponds to $z\,\rightarrow\,1$, $\zeta_{12}\,\rightarrow\,1$.}}. Any other direction at infinity corresponds to a singularity. The solution flows to the location of 3d $\mathcal{N}$ = 8 SYM phase, which is regular.

Lastly, when $\varphi(0)>\varphi_{cr}$, we start getting singular solutions. A singular solution is denoted by green lines in Figures 1 and 2. Singular solutions diverge at finite $r$.  On the contour plot in Figure 1, they flow to infinity which is not the 3d SYM phase.

We have also obtained analogous solutions around the $\mathcal{N}$ = 1, $SU(3)$ critical point. However, as they are similar to the solutions we obtained, we do not present them in the paper.

\vspace{1.1cm}

\section{The $G_2$-invariant Janus solutions}

The analysis of the $G_2$-invariant Janus solutions parallels the one in the previous section, and we will be brief on details. 

The scalar action for the $G_2$-invariant truncation, \cite{Guarino:2015qaa}, is obtained by \eqref{sub},
\begin{equation}
S\,=\,\frac{1}{16{\pi}G_4}{\int}d^4x\sqrt{-g}\left[R+14\frac{\partial_\mu{t}\partial^\mu\overline{t}}{(t-\overline{t})^2}-\mathcal{P}\,\right].
\end{equation}
The K\"ahler potential is
\begin{equation}
\mathcal{K}\,=\,-7\log\left[-i(t-\overline{t})\right]\,,
\end{equation}
and the holomorphic superpotential is
\begin{equation}
\mathcal{V}\,=\,14gt^3+2m\,.
\end{equation}
The scalar potential is obtained by
\begin{equation}
\mathcal{P}\,=\,2\left[-\frac{4}{7}(t-\bar{t})^2\left|\frac{\partial{W}}{\partial{t}}\right|^2-3W^2\right]\,.
\end{equation}
The scalar potential has $\mathcal{N}$ = 1, $G_2$-invariant critical point, and more non-supersymmetric critical points. The 3d $\mathcal{N}$ = 8 SYM phase is at $\chi\,=\,0$ and $\varphi\,\rightarrow\,-\,\infty$.

The supersymmetry equations for the complex scalar field,
\begin{equation}
t'\,=\,\left(-i \kappa\frac{e^{-A}}{l}-A'\right)\mathcal{K}^{t\bar{t}}\frac{2}{W}\frac{{\partial}W}{\partial\bar{t}}\,,
\end{equation}
and for the warp factor,
\begin{equation}
A'\,=\,\pm\sqrt{W^2-\frac{e^{-2A}}{l^2}}\,,
\end{equation}
are obtained where $\kappa\,=\,\pm1$. In terms of the real scalar fields, the supersymmetry equations  are
\begin{align}
\chi'+\frac{4}{7}e^{-2\varphi}\frac{A'}{W}\frac{{\partial}W}{\partial\chi}-\frac{4\kappa}{k}\frac{e^{-A-\varphi}}{l}\frac{1}{W}\frac{{\partial}W}{\partial\varphi}\,=&\,0\,, \notag \\
\varphi'+\frac{4}{7}\frac{A'}{W}\frac{{\partial}W}{\partial\varphi}+\frac{4\kappa}{k}\frac{e^{-A-\varphi}}{l}\frac{1}{W}\frac{{\partial}W}{\partial\chi}\,=&\,0\,.
\end{align}

Now we obtain numerical solutions in the same way we obtained the $SU(3)$-invariant solutions. As we will see in detail, there are three classes of solutions: regular Janus solutions, singular solutions, and as a special case of regular Janus solutions, there are Janus solutions flowing from a critical point. The solutions are depicted on contour plot of the superpotential in Figure 3.
\begin{figure}[h!]
\begin{center}
\includegraphics[width=3.0in]{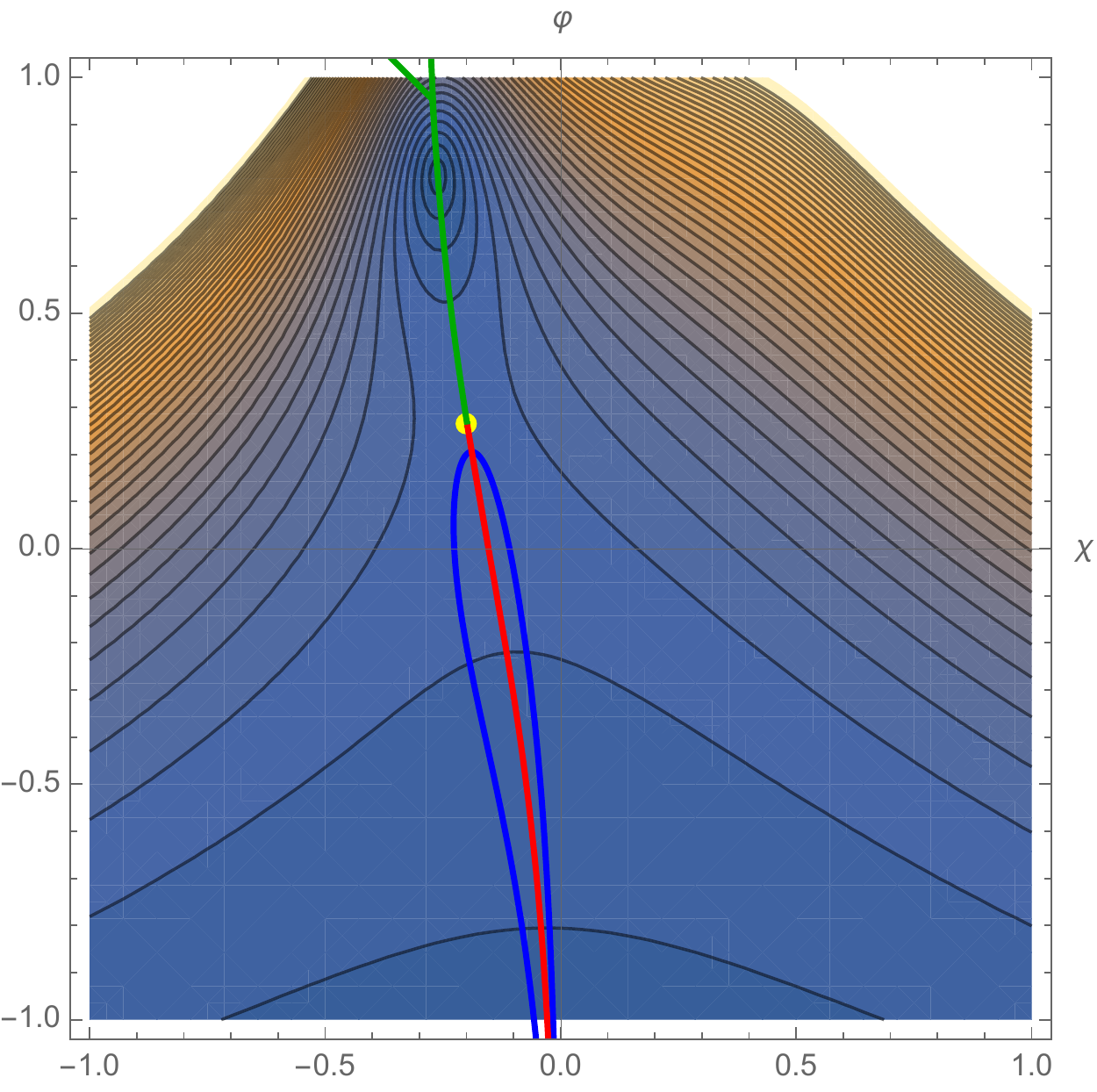}
\caption{{\it Contour plot of the $G_2$-invariant superpotential and flows. The $\mathcal{N}$ = 1, $G_2$ critical point is denoted by a yellow dot. The 3d $\mathcal{N}$ = 8 SYM phase is located at $\chi\,=\,0$ and $\varphi\,\rightarrow\,-\,\infty$.}}
\label{1}
\end{center}
\end{figure}

\begin{figure}[h!]
\begin{center}
\includegraphics[width=2.0in]{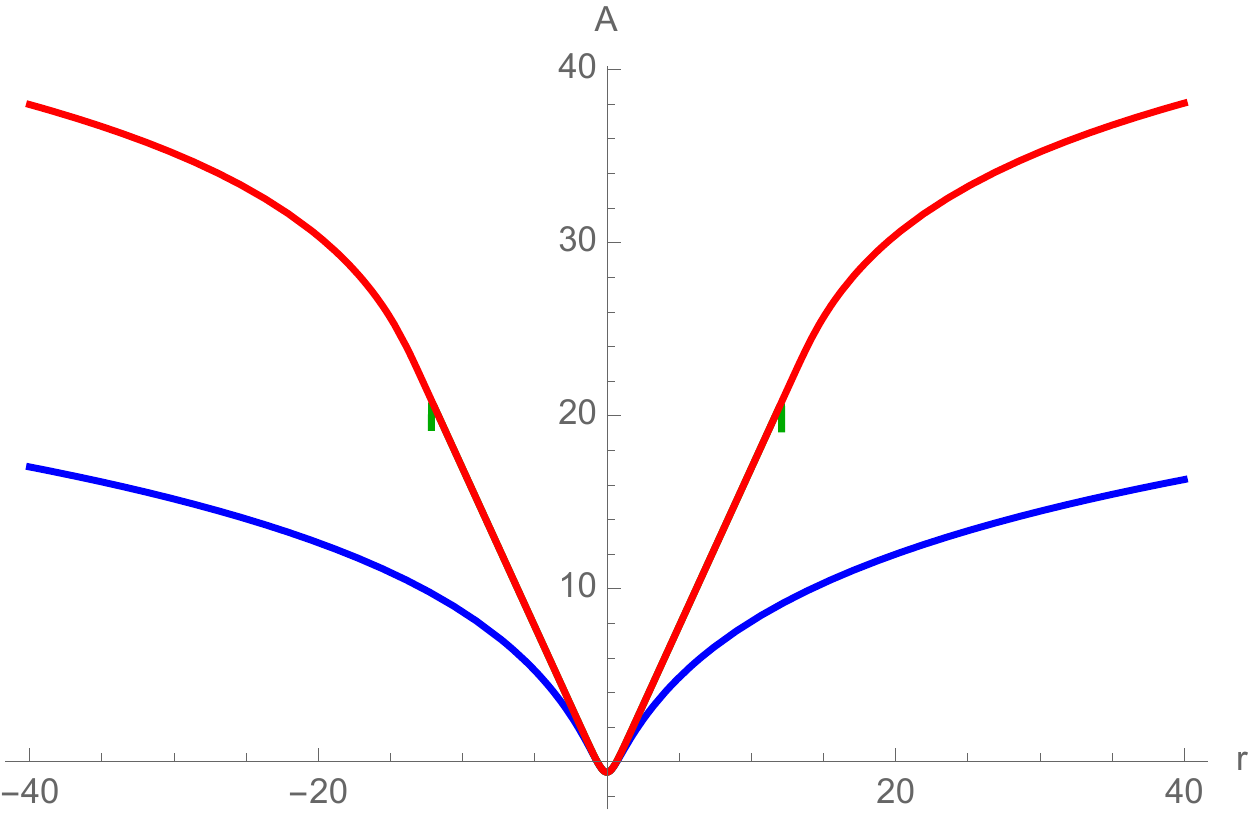} \qquad \includegraphics[width=2.0in]{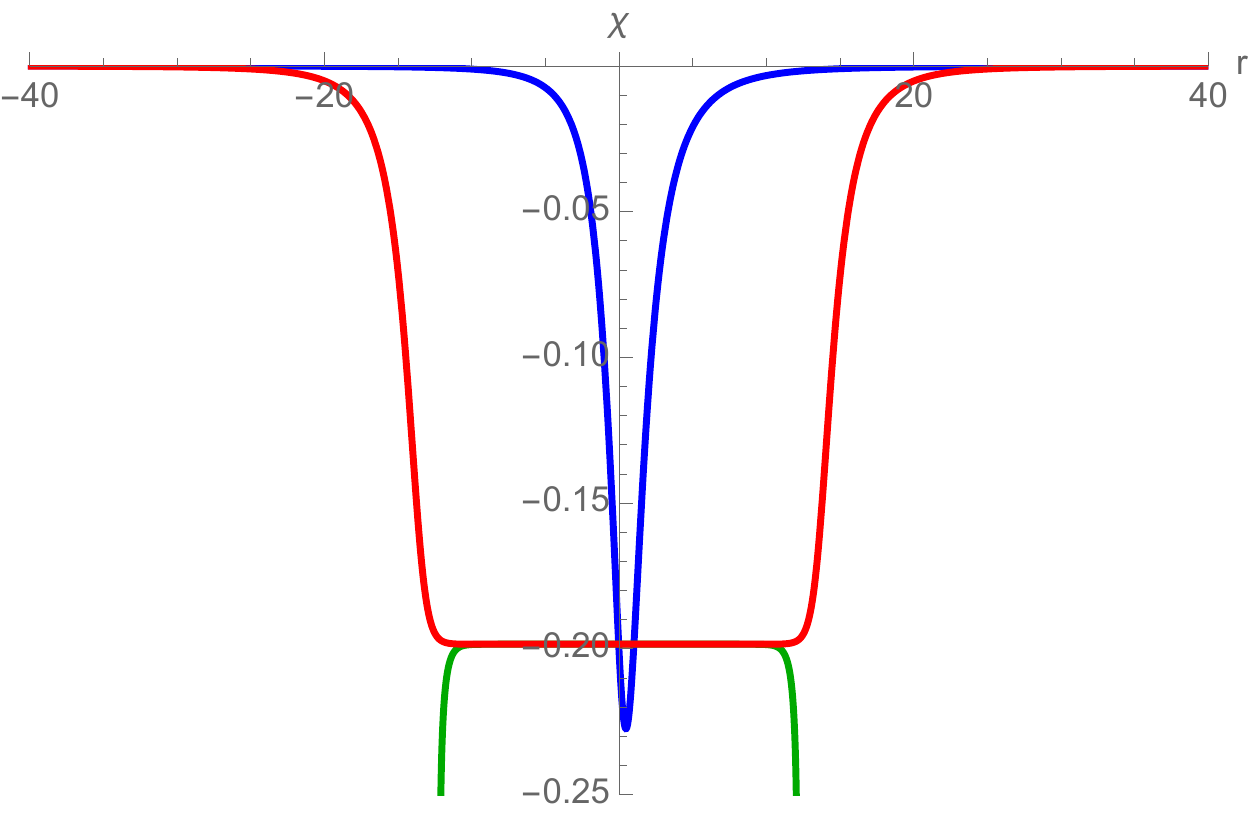} \\ \includegraphics[width=2.0in]{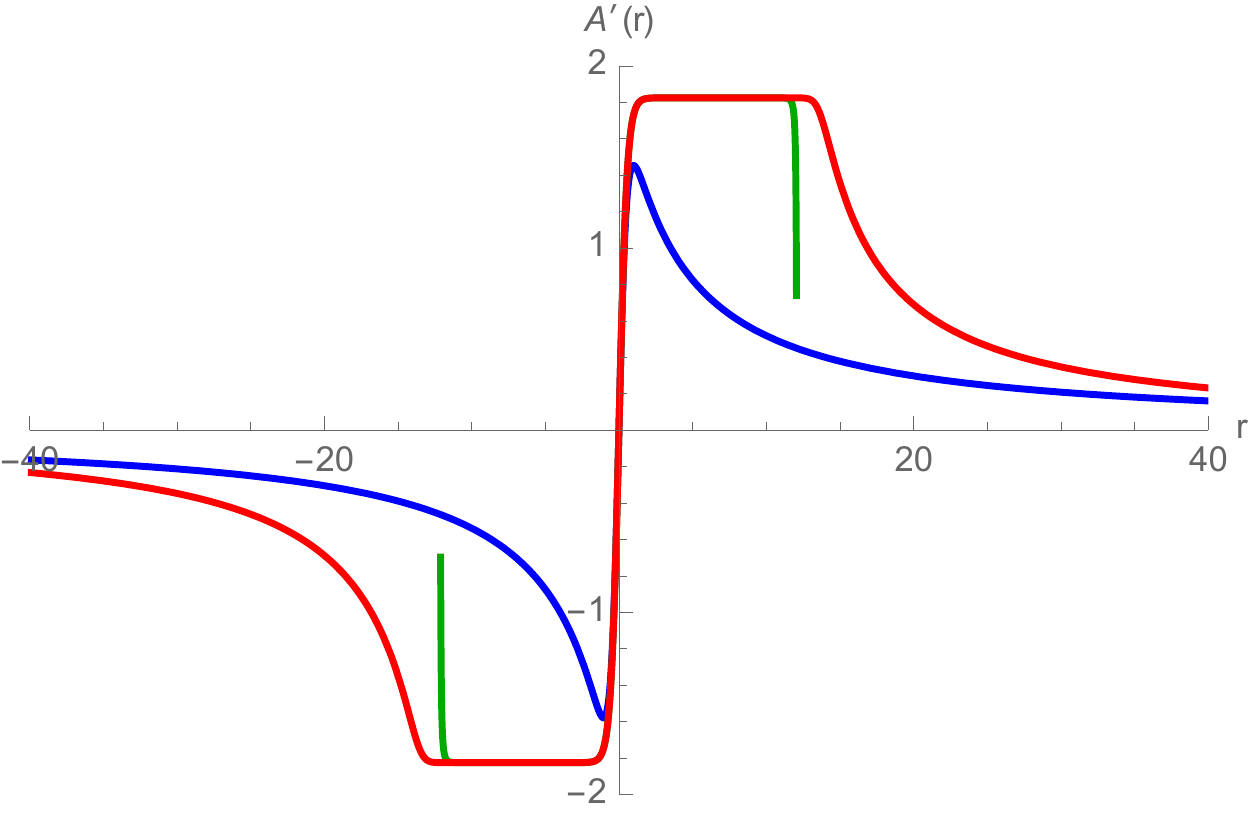} \qquad \includegraphics[width=2.0in]{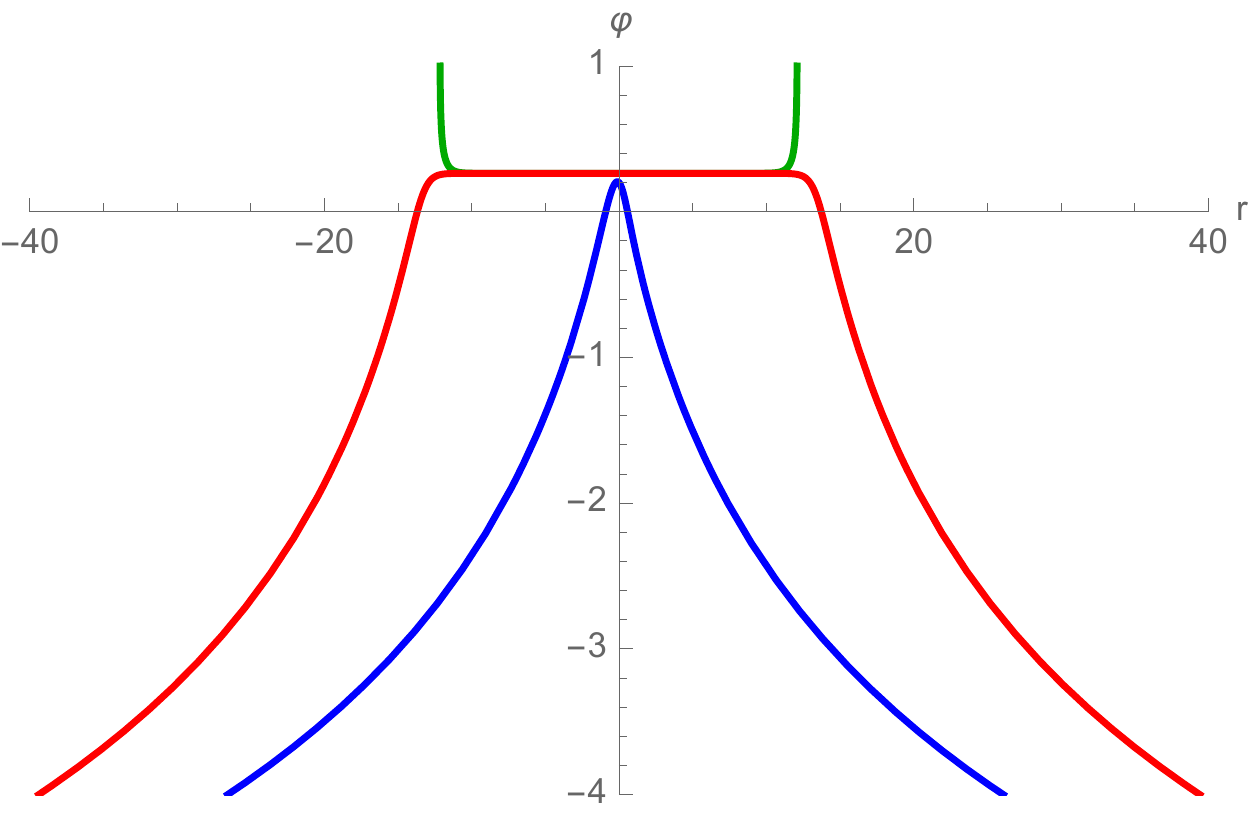}
\caption{{\it Three solutions with \{$\varphi(0)\,=\,0.2$, $\chi(0)\,=\,-0.2$\} (Blue), \{$\varphi(0)\,=\,\varphi_*+1.9\times10^{-16}$, $\chi(0)\,=\,\chi_*$\} (Red), and \{$\varphi(0)\,=\,\varphi_*+2.0\times10^{-16}$, $\chi(0)\,=\,\chi_*$\} (Green).}}
\label{1}
\end{center}
\end{figure}

First class of solutions are regular Janus solutions. Typical initial conditions give solutions attracted to 3d $\mathcal{N}$ = 8 SYM phase which is non-conformal. We plot a typical solution by blue lines in Figures 3 and 4.

The more interesting class of solutions arises as we approach the critical point. We fix $\chi(0)\,=\,\chi_*$, the value at the critical point. When $\varphi(0)<\varphi_{cr}$, we obtain the first class of solutions, $i.e.$ regular Janus solutions. We found the value of $\varphi_{cr}$ to be
\begin{equation}
\varphi_*+1.4\times10^{-16}\,\leq\,\varphi_{cr}\,\leq\,\varphi_*+1.9\times10^{-16}\,.
\end{equation}
When we fine-tune $\varphi(0)=\varphi_{cr}$, numerical solution starts approaching the critical point, stays at the critical point for a while, and then move away to the 3d SYM phase. While the solution is staying at the critical point, the warp factor exhibits the $AdS$-behavior with a constant slope. The warp factor is given by
\begin{equation}
A(r)\,\sim\,W_*r,
\end{equation}
where the slope, $W_*$, is the value of the superpotential at the critical point. We plot a typical critical-point solution by red lines in Figures 3 and 4. On the contour plot in Figure 3, this solution takes the steepest descent from the $G_2$ critical point to the SYM phase.

Lastly, when $\varphi(0)>\varphi_{cr}$, we start getting singular solutions. A singular solution is denoted by green lines in Figures 3 and 4.

\section{Conclusions}

In this paper, we numerically obtained supersymmetric Janus solutions in dyonic $ISO(7)$-gauged $\mathcal{N}\,=\,8$ supergravity. Unlike the Janus solutions known in four-and five-dimensional gauged supergravity, our Janus solutions are mostly attracted to the non-conformal 3d $\mathcal {N}$ = 8 SYM phase, which is the worldvolume theory on D2-branes. We also discovered a number of solutions which stay at a critical point for a while and then move to the SYM phase.

In dyonic $ISO(7)$-gauged $\mathcal{N}\,=\,8$ supergravity, there is also the $\mathcal{N}$ = 3, $SO(4)$ critical point. Interestringly, this critical point is non-supersymmetric in the $SO(4)$-invariant truncation we consider. It is because the $SO(4)$-invariant gravitino becomes massive at the critical point and breaks supersymmetry. If we consider the full $\mathcal{N}$ = 8 supergravity, there are three gravitinos outside the $SO(4)$-invariant sector, and hence the critical point is $\mathcal{N}$ = 3 supersymmetric, \cite{Guarino:2015qaa, Guarino:2016ynd}. Therefore, this critical point is not captured by the superpotential or the supersymmetry equations of our truncations. For this reason, we could not study Janus solutions of this critical point. It will be interesting to study Janus solutions of this critical point by employing a larger truncation preserving supersymmetry of this critical point. We present $\mathcal{N}$ = 1, $SO(4)$-invariant truncation in appendix B for completeness.

\bigskip
\leftline{\bf Acknowledgements}
We are grateful to Hyojoong Kim and Krzysztof Pilch for helpful discussions and communications. We would like to thank Adolfo Guarino for explaining his works, and Hyojoong Kim and Nakwoo Kim for collaboration on related subjects. We also would like to thank Adolfo Guarino and Krzysztof Pilch for comments on the preprint. This work was supported by the National Research Foundation of Korea under the grant NRF-2017R1D1A1B03034576.

\appendix
\section{The equations of motion}
\renewcommand{\theequation}{A.\arabic{equation}}
\setcounter{equation}{0} 

We present the equations of motion in this appendix. For the $SU(3)$-invariant truncation, the equations of motion are given by
\begin{align}
2A''+3A'A'+\frac{1}{l^2}e^{-2A}\,=\,-\left(\frac{3}{2}\varphi'\varphi'+\frac{3}{2}e^{2\varphi}\chi'\chi'+2\phi'\phi'+2e^{2\phi}\rho'\rho'-\mathcal{P}\right)\,, \notag \\
3A'A'+\frac{3}{l^2}e^{-2A}\,=\,\left(\frac{3}{2}\varphi'\varphi'+\frac{3}{2}e^{2\varphi}\chi'\chi'+2\phi'\phi'+2e^{2\phi}\rho'\rho'+\mathcal{P}\right)\,,
\end{align}
\begin{align}
\frac{3}{2}e^{-3A}\partial_\mu(e^{3A}g^{\mu\nu}\partial_\nu\varphi)-\frac{3}{2}e^{2\varphi}g^{\mu\nu}\partial_\mu\chi\partial_\nu\chi-\frac{1}{2}\frac{\partial\mathcal{P}}{\partial\varphi}\,=\,0\,, \notag \\
\frac{3}{2}e^{-3A}\partial_\mu(e^{3A+2\varphi}g^{\mu\nu}\partial_\nu\chi)-\frac{1}{2}\frac{\partial\mathcal{P}}{\partial\chi}\,=\,0\,, \notag \\
2e^{-3A}\partial_\mu(e^{3A}g^{\mu\nu}\partial_\nu\phi)-2e^{2\phi}g^{\mu\nu}\partial_\mu\phi\partial_\nu\phi-\frac{1}{2}\frac{\partial\mathcal{P}}{\partial\phi}\,=\,0\,, \notag \\
2e^{-3A}\partial_\mu(e^{3A+2\phi}g^{\mu\nu}\partial_\nu\rho)-\frac{1}{2}\frac{\partial\mathcal{P}}{\partial\rho}\,=\,0\,.
\end{align}
For the $G_2$-invariant truncation, the equations of motion are given by
\begin{align}
2A''+3A'A'+\frac{1}{l^2}e^{-2A}\,=\,-\left(\frac{7}{2}\varphi'\varphi'+\frac{7}{2}e^{2\varphi}\chi'\chi'-\mathcal{P}\right)\,, \notag \\
3A'A'+\frac{3}{l^2}e^{-2A}\,=\,\left(\frac{7}{2}\varphi'\varphi'+\frac{7}{2}e^{2\varphi}\chi'\chi'+\mathcal{P}\right)\,,
\end{align}
\begin{align}
\frac{7}{2}e^{-3A}\partial_\mu(e^{3A}g^{\mu\nu}\partial_\nu\varphi)-\frac{7}{2}e^{2\varphi}g^{\mu\nu}\partial_\mu\chi\partial_\nu\chi-\frac{1}{2}\frac{\partial\mathcal{P}}{\partial\varphi}\,=\,0\,, \notag \\
\frac{7}{2}e^{-3A}\partial_\mu(e^{3A+2\varphi}g^{\mu\nu}\partial_\nu\chi)-\frac{1}{2}\frac{\partial\mathcal{P}}{\partial\chi}\,=\,0\,.
\end{align}

\section{The $SO(4)$-invariant truncation}
\renewcommand{\theequation}{B.\arabic{equation}}
\setcounter{equation}{0} 

The scalar action for the $SO(4)$-invariant truncation, \cite{Guarino:2015qaa, Guarino:2016ynd}, is obtained by \eqref{sub},
\begin{equation}
S\,=\,\frac{1}{16{\pi}G_4}{\int}d^4x\sqrt{-g}\left[R+12\frac{\partial_\mu{t}\partial^\mu\overline{t}}{(t-\overline{t})^2}+2\frac{\partial_\mu{u}\partial^\mu\overline{u}}{(u-\overline{u})^2}-\mathcal{P}\,\right].
\end{equation}
The K\"ahler potential is
\begin{equation}
\mathcal{K}\,=\,-6\log\left[-i(t-\overline{t})\right]-\log\left[-i(u-\overline{u})\right]\,,
\end{equation}
and the holomorphic superpotential is
\begin{equation}
\mathcal{V}\,=\,2g\left(4t^2+3t^2u\right)+2m\,.
\end{equation}
The scalar potential is obtained from{\footnote{We have corrected some misprints in appendix C.1 of \cite{Guarino:2016ynd}.}}
\begin{equation}
\mathcal{P}\,=\,2\left[-\frac{2}{3}(t-\bar{t})^2\left|\frac{\partial{W}}{\partial{t}}\right|^2-4(u-\bar{u})^2\left|\frac{\partial{W}}{\partial{u}}\right|^2-3W^2\right]\,.
\end{equation}
The scalar potential has $\mathcal{N}$ = 1 $G_2$- and $\mathcal{N}$ = 3 $SO(4)$-invariant critical points, and more non-supersymmetric critical points.



\end{document}